\documentclass[pra,aps,twocolumn,superscriptaddress]{revtex4-2}
\usepackage{graphicx} 
\usepackage{color}
\usepackage{epsfig}
\usepackage{dsfont}
\usepackage{hyperref}
\usepackage{amsmath,amsfonts}

\begin{document}
\title{Multipartite Entanglement versus Multiparticle Entanglement}
\author{Marcin Wie\'sniak}
\affiliation{Instiute o Theoretical Physics and Astrophysics, Faculty of Mathematics, Physics, and Astrophysics,\\ University of Gda\'sk, ul Wita Stwosza 57, 80-309 Gda\'nsk, Poland.}
\email{marcin.wiesniak@ug.edu.pl}

\begin{abstract}
    Entanglement is defined as presence of quantum correlations beyond those achieved by local action and classical communication. To identify its presence in a generic state, one can, for example, check for existence of a decomposition of separable states. A natural extension is a genuine multipartite entanglement (GME), understood as nonexistenence of a decomposition into biseparable states (later called biseparable decomposition, BD). In this contribution we revisit activation of GME. We discuss few examples of states, which are decomposable into a mixture of biproduct states. However, after merging two copies of these states, we certify nonexistence of BD with witness operators. This seems to challenge our understanding of GME as a separate resource. It turns out that it requires a careful consideration of the physical context. We stress that activation of GME from  multiple copies of GME-free states necessarily involves entangling operations.
\end{abstract}
\date{\today}
\newtheorem{algorithm}{Algorithm}
\newtheorem{definition}{Definition}
\newcommand{\bra}[1]{\left\langle#1\right|}
\newcommand{\ket}[1]{\left|#1\right\rangle}
\newcommand{\abs}[1]{\left|#1\right|}
\newcommand{\mean}[1]{\left\langle #1\right\rangle}
\newcommand{\braket}[2]{\left\langle{#1}|{#2}\right\rangle} 
\newcommand{\commt}[2]{\left[{#1},{#2}\right]}
\newcommand{\tr}[1]{\mbox{Tr}{#1}}
\maketitle

\section{Introduction}
Quantum mechanics diverges from classical physics with including the superposition principle. For an individual quantum system this means a possibility to experience more than one classically understood history. An even more striking effect is entanglement, which allows for a superposition of two, possibly spatially separated, subsystems to be in a superposition \cite{schrodinger}. That is to say, either subsystem's properties are described only in reference to the other, e.g., spins of both particles are always anticorrelated. 

{
With the rise of interest in the Bell's theorem \cite{bell}, entanglement turned out to a be an information processing resource useful in, for example, quantum cryptography \cite{ekert}. On the other hand, researchers found it to be surprisingly complex to characterize. A pure state of two systems is entangled iff it is not represented as a product of two local states. Entanglement of a mixed state is defined by nonexistence of a separable decomposition,
\begin{definition}
\label{def1}
$\rho$ is entangled iff  
\begin{eqnarray}
    \label{ENTDEF}
    &&\rho\neq\sum_{j}p_j\Pi(\ket{\psi_j}\otimes\ket{\phi_j}),\nonumber\\
    &&\sum_jp_j=1,p_j\geq 0,\nonumber\\
    &&\Pi(\ket{\psi})=\ket{\psi}\bra{\psi}.
\end{eqnarray}
\end{definition}

As it is not feasible to consider all possible decompositions, it has been not considered highly practical to verify presence of entanglement with it and many additional criteria for entanglement were formulated. A negative eigenvalue of the state under partial transposition was a strong indicator of the presence of quantum correlations \cite{peres,horodecki1996}, but not a universal one, as PPT entangled states exist \cite{horodecki1997}. Any form of entanglement can be detected by an application of a positive, but not a completely positive map on the state \cite{horodecki1996}, but for a given state it is unfeasible to check all possible positive maps. By Jamio\l{}kowski-Choi isomorphism \cite{jamiolkowski,choi} any such map corresponds to an entanglement witness operator \cite{terhal}, which can  attain positive (by a convention taken here) mean value only for a certain class of entangled states. However, until recently we have lacked an efficient way to find the witness operator.}

The entanglement detection problem affects the efforts in quantifying this resource. Obviously, quantum correlations are a quantitative rather than a qualitative feature, and there has been a number of propositions for their proper measures, e.g., \cite{bennett1996,vedral,eisert2003,chen,hiesmayr}. But, again, so far, any formulation of such a measure would be a compromise between some reasonable requirements (discussed below), mainly sacrificing its ability to be efficiently computed.

Even with our incomplete understanding of bipartite entanglement, it is a natural next step to discuss genuinely multipartite entanglement (GME). For example, multipartite quantum correlations have been shown to lead to a stronger all-versus-nothing conflict between quantum mechanics and local hidden variable models \cite{greenberger} than the original Bell inequality for two qubits \cite{bell}. Correlations between multiple system may also be necessary to reach the ground state energy in certain interacting spin lattices \cite{guehne}, or to generate word states in an error-correcting code \cite{steane}.

It is thus tempting to generalize definition (\ref{def1}) to
\begin{definition}
\label{def2}
$\rho$ is genuinely multipartite entangled iff
\begin{eqnarray}
    \label{MENTDEF}
    &&\Leftrightarrow \rho\neq\sum_{\mathcal{A}\subset \Omega}\sum_{j_{\mathcal{A}}}p_{j_\mathcal{A}}\Pi(\ket{\psi_{j_\mathcal{A}}}_\mathcal{A}\otimes\ket{\phi_{j_\mathcal{A}}}_{\mathcal{A}^C}),\nonumber\\
    &&\sum_{\mathcal{A}\subset \Omega}\sum_{j_{\mathcal{A}}}p_{j,\mathcal{A}}=1,p_{j,\mathcal{A}}\geq 0.
\end{eqnarray}
\end{definition}

Here, $\Omega$ is the set of all parties describing state $\rho$, $\mathcal{A}$ goes over all its proper subsets and $\mathcal{A}^C$ is the complement of $\mathcal{A}$ in $\Omega$. The right-hand side of the first line of definition (\ref{MENTDEF}) will be called the biseparable decomposition and abbreviated as BD.

Recently, Yamasaki {\em et al.} \cite{yamasaki} have discussed the phenomenon that for multiple copies of certain states, BD breaks down. The Authors of Ref. \cite{yamasaki} dub this effect {\em activation of genuine multipartite entanglement}. In this contribution, we investigate the relation of this effect to the requirements for entanglement measures. As we shall see, the usual interpretation leads to an inevitable conclusion that genuine multipartite entanglement cannot be a proper resource as it can be produced by either simply taking multiple copies of the same GME-free state, or by applying supposedly local operations. We also investigate the third logical possibility, namely that these operations should not be considered local. 

In this contribution we investigate few most basic cases, in which, according to definition (\ref{def2}) and Ref. \cite{yamasaki}, genuine multipartite entanglement is activated with two copies. This is confirmed by a witness operator. However, a single copy cannot be multipartite entangled by construction. While this is well accepted by Yamasaki {\em et al.}, we believe that this point of view calls for a deepened discussion. While the question of number of copies and the nature of operations needed to create GME is relevant, it needs to be put in context of operational feasibility. We shall discuss an alternative formulation of the multipartite entanglement problem, which is free of these difficulties.
\section{Requirements for bipartite and multipartite entanglement measures}
It is not surprising that entanglement has been considered a resource, both in the sense of its potential advantage in communication and computation tasks, and of the resource theory, where we distinguish states and operations that are free or rich in the resource. Separable states are free, as all of them can be obtain by LOCC.  Likewise, local operations, possibly coordinated among participants, cannot be used to create or increase entanglement, so they are considered free. This opens a whole theory of transformation between different states. Too understand transition between which states are possible, and at what rates, are possible, we need some means to quantify the amount of our resource. The basic requirements for such a measure $E$ in the bipartite case are \cite{bengtsson}: 
\begin{enumerate}
    \item{{\em Discriminance:} $E(\rho)=0$ if $\rho$ is not entangled, i.e. there exist probabilities $p_j$ and product state $\ket{\psi_j}\otimes\ket{\varphi_j}$, such that $\rho=\sum_jp_j\Pi(\ket{\psi_j}\otimes\ket{\varphi_j})$.}
    \item{{\em Monotonicity:} $E(\rho)$ cannot increase under stochastic local actions and classical communication (SLOCC).}
    \item{{\em Convexity:} $E(p\sigma+(1-p)\rho)\leq p E(\sigma)+(1-p)E(\rho)$, where $0\leq p\leq 1$}
    \item{{\em Asymptotic continuity:} Let $\rho_m$ and $\sigma_m$ denote series of state acting on Hilbert spaces $(\mathcal{H}_{d_1}\otimes\mathcal{H}_{d_2})^{\otimes m}$. Then
    $||\rho_m-\sigma_m||_1\rightarrow 0\Rightarrow\frac{E(\rho_m)-E(\sigma_m)}{m}\rightarrow 0$.}
    \item{{\em Normalization:} $E(\Pi(\ket{\Psi^+}))=1$, where $\ket{\Psi^+}=(\ket{00}+\ket{11})/\sqrt{2}$.}
    \item{{\em Additivity:} $E(\rho\otimes\sigma)=E(\rho)+E(\sigma)$.}
    \item{{\em Computability:} There is an efficient way to compute $E(\rho)$ for any $\rho$.}
\end{enumerate}

Jointly, these conditions up to date have prevented us from formulating an universal measures of bipartite entanglement. Some candidate figures of merit were proposed and proven to satisfy most of the requirements, but largely fail to be computable. Examples include entanglement cost, distillable entanglement \cite{bennett1996}, or measures based on extensions to pure states \cite{vedral}. Therefore, some requirements were modified. For example, asymptotic continuity can be efficiently replaced by regular additivity, $\rho_m\rightarrow_{m\rightarrow\infty}\rho\Rightarrow E(\rho_m)\rightarrow_{m\rightarrow\infty}E(\rho)$, which is much easier to prove. Monotonicity can be replaced by weaker {\em monotonicity under deterministic LOCC}, and in this contribution we will consider an even weaker requirement {\em no production of (genuinely multipartite) entanglement} under SLOCC. Likewise, additivity can be relaxed to {\em extensivity} stating that $E(\rho^{\otimes n})=n E(\rho)$, but for here it will be sufficient to consider even weaker statement, {\em no production} under copying the state. Negativity-related entanglement indicators \cite{zyczkowski,eisert1999} are straight-forwardly computable, but they largely fail the discriminance criterion due existence of bound entangled states.

On the other hand, one can formulate entanglement indicators and measures based on various distance functions based on relative entropy \cite{henderson}, Bures \cite{vedral}, trace \cite{eisert2003}, and Hilbert-Schmidt (HS) distances \cite{witte}. Their main advantages are discriminance, and, in particular in case of the Hilbert-Schmidt distance, convexity. While the exact computation of distance between a given state and a convex set remains difficult, a recent adaptation of the Gilbert \cite{gilbert} algorithm \cite{pandya} allowed to efficiently give both a precise upper and lower bounds on HS distance to the closest separable state. More recently, similar techniques were applied to the Bures distance \cite{hu}.

Interestringly, a brief review on multipartite entanglement measure \cite{ma2023} follows Ref. \cite{ma2011} in not mentioning additivity or extensivity as a minimal requirement for a measure of genuine multipartite entanglement. However, Refs. \cite{mintert,hiesmayr}, to which Ref. \cite{ma2011} redirects the Reader mention these properties. Also, it is worth noticing that Ref. \cite{bengtsson} quotes both subadditivity and superadditivity as possible relaxations of requirement 6.
\section{Gilbert algorithm}
We now briefly recall the Gilbert algorithm, which yields an approximation of the closest separable state to a given state, with respect to Hilbert-Schmidt distance, $D_{HS}(A,B)=\sqrt{{\text{Tr}}(A-B)(A-B)^\dagger}$. 
\begin{algorithm} (bipartite case):
\label{Gilbert}
\\
{\em Input:} test state $\rho$, initial separable state $\rho'$.\\
{\em Output:} approximation of CSS $\rho'$, list of squared distances to subsequent CSS approximations $l$.
\begin{enumerate}
    {\item Take random pure state $\rho''=\ket{\varphi_A}\ket{\varphi_B}\bra{\varphi_A}\bra{\varphi_B}$, that will be referred to as a trial state.}
    {\item If the preselection criterion, ${\text{Tr}}(\rho-\rho')(\rho''-\rho')>0$ is not met, go to step 1 or abort if the HALT condition is satisfied.}
    {\item Maximize $\text{Tr}(\rho-\rho')(\rho''-\rho')$ with local unitary transformations.}
    {\item Update $\rho_1\leftarrow p \rho'+(1-p)\rho''$ for $p=\underset{0\leq p'\leq 1}{{\text Argmax}}D_{HS}(\rho,p' \rho'+(1-p')\rho'')$.}
    {\item Every 50 corrections append $D^2_{HS}(\rho,\rho')$ to $l$.}
    {\item If the HALT condition is not met, go to step 1, otherwise quit.}
\end{enumerate}
\end{algorithm}

The generalization of the algorithm to finding the existence of BD is straight-forward, one just replaces a product state in step 1 with a state which is a biproduct with respect to different partition. After suuficiently many repetiotions of the algorithm we obtain three quantities informing us about the (multipartite) entanglement content. First is the last distance found after, e.g., a given number of corrections, $D_\text{Last}=\sqrt{\text{Tr}(\rho-\rho')^2}$. The second is the witness distance. The witness operator is defined as \cite{brandao}:
\begin{eqnarray}
    W=&&\frac{\rho-\rho'-\lambda\mathds{1}}{D_\text{Last}},\nonumber\\
    \lambda=&&\max_{\ket{\psi}{\text{ is bipartite product}}}\bra{\psi}(\rho-\rho')\ket{\psi}. 
\end{eqnarray}
The witness distance is defined as $D_{Wit}=\text{max}(0, \text{Tr}\rho W)$.

It is also useful to estimate the distance from the decay of $D^2_{\text{Last}}$, which is stored in the list $l$. First we create list $\tilde{l}$ by rejecting the first one third of the entries. Second, we shift $\tilde{l}$ by $a$ an inverse the entries. With $c$ being the list of consecutive numbers of correction and $\overline{x}$ denoting the mean value of list $x$, the squared estimated distance is defined as
\begin{eqnarray}
    D_{\text{Last}}^2&&=\underset{0\leq a\leq D_{\text{Last}}}{\text{Argmax}}(R(c,1/(\tilde{l}-a))),\nonumber\\
    R(x,y)=&&\frac{\overline{xy}-\overline{x}\cdot\overline{y}}{\sqrt{\left(\overline{x^2}-(\overline{x})^2\right)\left(\overline{y^2}-(\overline{y})^2\right)}}.
\end{eqnarray}
Here, $\overline{x}$ denotes the average value of the list, and the product of two lists is entry-wise.

Importantly, $D_{\text{Last}}$ is the upper bound on the actual distance, while $D_{\text{Wit}}$ is a lower bound. In particular, finding $D_{\text{Wit}}>0$ certifies nonexistence of BD. On the other hand, the Algorithm cannot reach $D_{\text{Last}}=0$ but for a large class of states it can made arbitrary low with a sufficiently long runtime, leading us to a strong belief that the studied state is indeed separable. Thus the Gilbert algorithm provides a high level of discriminance and computability of the Hilbert-Schmidt distance. However, it is known to that it violates contractiveness under LOCC.

\section{Early realizations of genuine multipartite entangled states}
In this section we will briefly recall two pioneering quantum optical experiments, which led to observation of multipartite entanglement. They will turn out both conceptually relevant to the examples of GME states discussed below and helpful in considering their feasibility.

These realizations were based spontaneous parametric down-conversion (SPDC) \cite{louisell}. In short, it is a process, in which inside a nonlinear crystal a high-frequency (pump) photon is converted into a highly correlated pair of low-frequency photons. Ultrastrong, strictly quantum correlations between the output (signal and idler) photons include the frequencies summing to the one of the pump photon, times of creation being equal, their positions being symmetrical with respect to a fixed axis, and, in type-II SPDC, polarization. For example, the interaction Hamiltonian between the pump beam and the output field can have term
\begin{eqnarray}
    H_{\text{Int}}=&&i(za_{p,H}a^\dagger_{c,H}a^\dagger_{d,H}-z^*a^\dagger_{p,H}a_{c,H}a_{d,H}),
\end{eqnarray}
with $z$ being a complex amplitude, $a_{x,Y}$ -- the annihilation operator with mode $x$ ($p$ for pump, $c$ for signal, $d$ for ilder) and polarizations $X=H,V$.

In the first approximation, the output of SPDC, i.e. the (four-mode) squeezed vacuum can be written as $\alpha\ket{\Omega}+\beta(\ket{00}+\ket{11})/\sqrt{2}=\alpha\ket{\Omega}+\beta\ket{\Psi^+}$, where $\alpha$ and $\beta$ are some coefficients, $\ket{\Omega}$ denotes the vacuum in output modes, and $\ket{00}$ ($\ket{11})$ denotes two photons with, say, horizontal (vertical) polarization. Discarding the vacuum part by postselection and taking two copies of the reminder we get $\ket{\Psi^+}_{AB}\ket{\Psi^+}_{CD}$. After crossing signals in modes $B$ and $C$ on a polarizing beam splitter (PBS) and demanding that one photon goes to each observer, we get the GHZ state \cite{pan},
\begin{eqnarray}
\ket{GHZ_4}=&&(\ket{0000}_{ABCD}+\ket{1111}_{ABCD}),\nonumber\\
\ket{0}\equiv&&\left(\begin{array}{c}1\\0\end{array}\right),\quad\ket{1}\equiv\left(\begin{array}{c}0\\1\end{array}\right),
\end{eqnarray}
which is a seminal example of a state with genuine four-partite entanglement, namely the GHZ state \cite{greenberger}. In the same fashion and by measuring one photon, e.g., in the basis of $L/R$ polarization, we can obtain the GHZ state for any number of qubits. Note that the requiring that both photons reaching PBS have the same polarization is a projection onto two-dimensional space.

The other example to be recalled here is obtained when a single SPDC source emits two pairs at once. In the second quantization, we write it as proportional to 
$(\hat{a}_H^\dagger\hat{c}_H^\dagger+\hat{a}_V^\dagger\hat{c}_V^\dagger)^2\ket{\Omega}$. Next, output channels $a$ and $c$ are split with a non-polarizing beam splitter into modes $a$, $b$, and $c$, $d$, respectively. With a post-selection condition that one photon propagates to one output channel, the state is \cite{gaertner}
\begin{eqnarray}
&&\ket{\Psi_4}\nonumber\\
\propto&&(2\hat{a}_H^\dagger\hat{b}_H^\dagger\hat{c}_H^\dagger\hat{d}_H^\dagger+2\hat{a}_V^\dagger\hat{b}_V^\dagger\hat{c}_V^\dagger\hat{d}_V^\dagger+\hat{a}_H^\dagger\hat{b}_V^\dagger\hat{c}_H^\dagger\hat{d}_V^\dagger\nonumber\\+&&\hat{a}_H^\dagger\hat{b}_V^\dagger\hat{c}_V^\dagger\hat{d}_H^\dagger+\hat{a}_V^\dagger\hat{b}_H^\dagger\hat{c}_H^\dagger\hat{d}_V^\dagger+\hat{a}_V^\dagger\hat{b}_H^\dagger\hat{c}_V^\dagger\hat{d}_H^\dagger)\ket{\Omega}\nonumber\\
\propto&&\frac{1}{\sqrt{3}}\left(\ket{0000}_{ABCD}+\ket{1111}_{ABCD}\right.\nonumber\\
&&+\left.\ket{\Psi^+}_{AC}\otimes\ket{\Psi^+}_{BD}\right).
\end{eqnarray}
This state is equivalent, up to local unitary transformations, to the four-qubit singlet state, in which two pairs spins-$1/2$ jointly create two spins-$1$, which then nullify each other. We shall stress that the key ingridients are the beam splitters and postselection. While the description of this scheme requires the second quantization, in which we lack a proper definition of entanglement, combination of these two elements represent an entangling action on two particles.
\section{Multiple copies of some states loose BD}
First, let us consider a simple case of a pure state. Let $B$ simultaneously share $\ket{\Psi^+}$ with both $A$ and $C$, which can be written as
\begin{eqnarray}
\label{varphi4}
&&\ket{\varphi_4}\nonumber\\
=&&\ket{\Psi^+}_{AB_1}\ket{\Psi^+}_{B_2C}\nonumber\\
\approx&&\frac{1}{2}\sum_{i,j=0}^1\ket{i}_A\ket{ij}_B\ket{j}_C.
\end{eqnarray}
This state does not have a biseparable decomposition. In fact, consider tripartite negativity \cite{sabin},
\begin{eqnarray}
    \mathcal{N}_{ABC}=&&(\mathcal{N}_{A|BC}\mathcal{N}_{B|AC}\mathcal{N}_{C|AB})^{\frac{1}{3}},\nonumber\\
    \mathcal{N}_{I|JK}=&&-2\sum_{\epsilon_i<0}\epsilon_i(\rho^{\Gamma_I}),
\end{eqnarray}
where $\rho^{\Gamma_I}$ is the state partially transposed with respect to subsystem $I$ and $\epsilon_i(\rho^{\Gamma_I})$s are its eigenvalues. The negativity of $\ket{\varphi_4}\bra{\varphi_4}$ transposed with respect to $A$ and $C$ is 1,  while for $B$ it equals 3, giving $\mathcal{N}_{ABC}=3^{\frac{1}{3}}\approx 1.4422$. 

Notice that the state $\ket{\Psi^+}_{AB_1}\ket{\Psi^+}_{B_2C}$ is obtainable both from two copies of GHZ states and two copies of the three-party W state,
\begin{equation}
    \ket{W}=\frac{1}{\sqrt{3}}(\ket{001}+\ket{010}+\ket{001}).
\end{equation}
and it can be transformed into a single copy of either of them. To pass from two GHZ states to a W state, $A$ measures the second qubit in any basis in the $x-y$ plane, while $C$ does the same for the first qubit. Depending on their measurements and outcomes, the state of the remaining qubits is locally unitarily transformed to the state given in the middle line of $\ref{varphi4}$. Now, $B$ probabilisticly projects qubits with
\begin{equation}
    P_{B'}=\left(\begin{array}{cccc}0&\frac{1}{\sqrt{2}}&\frac{1}{\sqrt{2}}&0\\1&0&0&0\end{array}\right),
\end{equation}
and coherently attenuates the remaining qubit state $\ket{1}$ with 
\begin{equation}
    Attn=\left(\begin{array}{cc}1&0\\0&\frac{1}{\sqrt{2}}\end{array}\right).
\end{equation}
The overall probability of this conversion is $3/8$.

In the converse protocol, one copy of the w state is projected with $\ket{0_{A_2}}\bra{0_{A_2}}$, while the other is projected with$\ket{0_{C_1}}\bra{0_{C_1}}$. Consequently, we $\ket{\Psi^+}_{AB_1}\ket{\Psi^+}_{B_2C}$ and $B$ can project onto
\begin{equation}
\label{PB}
P_B=\left(\begin{array}{cccc}1&0&0&0\\0&0&0&1\end{array}\right)
\end{equation}
to obtain the GHZ state. The probability of of these projections being successful is $\frac{2}{9}$.

Now, let us discuss examples of states, two copies of which in this paradigm reveal genuine multipartite entanglement. Consider state
\begin{eqnarray}
\rho_1(\theta)=&&\frac{1}{2}\left(\Pi(\ket{\theta}_A)\otimes\Pi((\ket{00}_{BC}+\ket{11}_{BC})/\sqrt{2})\right.\nonumber\\
+&&\left.\Pi((\ket{00}_{AB}+\ket{11}_{AB})/\sqrt{2})\otimes\Pi(\ket{\theta}_C)\right)
\end{eqnarray}
with $\ket{\theta}=(\cos(\theta)\ket{0}+\sin(\theta)\ket{1})/\sqrt{2}$ and  $\Pi(\ket{\psi})=\ket{\psi}\bra{\psi}$.

By construction, $\rho_1$ satisfies condition (\ref{MENTDEF}). However, let us take two copies of $\rho_1$ shared between parties $A$, $B$, and $C$, so that each party holds two qubits identically correlated with the rest. While we do not have convincing enough arguments that $\rho_1(\theta)^{\otimes 2}$ are not decomposable into biseparable states, we let each observer perform a projection on a two-dimensional subspace. $B$ performs projection $P_B$
whereas $A$ and $C$ each project onto
\begin{equation}
P_{A/C}=\left(\begin{array}{cccc}\cos^2(\theta)&\cos(\theta)\sin(\theta)&\cos(\theta)\sin(\theta)&\sin^2(\theta)\\\frac{\sin(2\theta)}{\sqrt{2}}&-\frac{\cos(2\theta)}{\sqrt{2}}&-\frac{\cos(2\theta)}{\sqrt{2}}&\frac{\sin(2\theta)}{\sqrt{2}}\end{array}\right).
\end{equation}
The resulting state is hence
\begin{equation}
    \rho_3(\theta)=(P_{A/C}\otimes P_B\otimes P_{A/C})\rho_1(\theta)^{\otimes 2}(P_{A/C}\otimes P_B\otimes P_{A/C})^\dagger
\end{equation}
Figure 2 shows $D_{\text Last}$, $D_{\text Est}$, and $D_{\text Wit}$ in function of $\theta$ with Algorithm \ref{Gilbert} having performed between 1800 and 3500 corrections.
\begin{figure}[b]
    \centering
    \includegraphics[width=70mm]{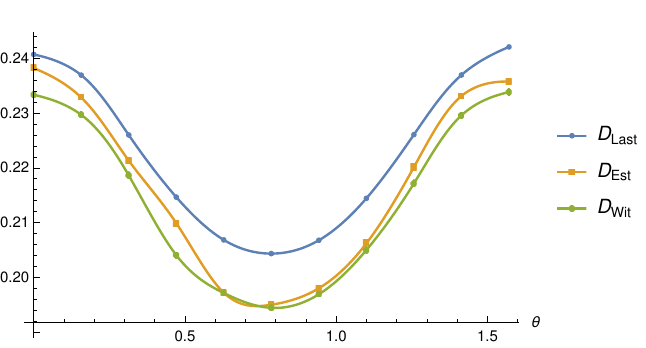}
    \caption{Hilbert-Schmidt distances $D_{\text{Last}}$,$D_{\text{Est}}$ and $D_{\text{Last}}$ found for $\rho_3(\theta)$.}
    \label{fig:enter-label}
\end{figure}

As seen in the figure, the resulting state has no BD for all values of $\theta$, as certified by $D_{\text Wit}$. In particular, for $\theta=0$ we have 
\begin{equation}
    \rho_3(0)=\frac{1}{9}\left(\begin{array}{cccccccc}
    8&0&0&0&0&0&0&2\\
    0&0&0&0&0&0&0&0\\
    0&0&0&0&0&0&0&0\\
    0&0&0&0&0&0&0&0\\
    0&0&0&0&0&0&0&0\\
    0&0&0&0&0&0&0&0\\
    0&0&0&0&0&0&0&0\\
    2&0&0&0&0&0&0&1\\
    \end{array}\right),
\end{equation}
for which the witness of decomposability into biseparable states is close to (up to a multiplicative constant)
\begin{eqnarray}
    &&W_3(0)=8\ket{GHZ_3}\bra{GHZ_3}-\mathds{1}_{8\times 8},\nonumber\\
    &&\max_{\ket{\psi}\text{ is biseparable}}\bra{\psi}W_3(0)\ket{\psi}=3,
\end{eqnarray}
where $\ket{GHZ_3}=\frac{1}{2}(\ket{000}+\ket{111})$. The witness indicates the distance from $\rho_3(0)$ and the set of biseparable states is $\sqrt{\frac{8}{63}}\approx 0.2376$, which is slightly larger than the witness distance found by the algorithm, $0.2333$

Another example we want to present here is a four-qubit state, composed of two pairs of Bell states,
\begin{eqnarray}
    &&\rho_2\nonumber\\
    =&&\frac{1}{2}\left(\ket{\Psi^+}_{AB}\bra{\Psi^+}_{AB}\otimes\ket{\Psi^+}_{CD}\bra{\Psi^+}_{CD}\right.\nonumber\\
    +&&\left.\ket{\Psi^+}_{AD}\bra{\Psi^+}_{AD}\otimes\ket{\Psi^+}_{BC}\bra{\Psi^+}_{BC}\right).\nonumber\\
\end{eqnarray}
Again, the parties share two-copies of the state and each performs a projection $P_B$ given by Eq. (\ref{PB}). The resulting state reads
\begin{widetext}
    \begin{equation}
        \rho_4=\frac{1}{12}\left(\begin{array}{cccccccccccccccc}
 4 & 0 & 0 & 1 & 0 & 0 & 1 & 0 & 0 & 1 & 0 & 0 & 1 & 0 & 0 & 4 \\
 0 & 0 & 0 & 0 & 0 & 0 & 0 & 0 & 0 & 0 & 0 & 0 & 0 & 0 & 0 & 0 \\
 0 & 0 & 0 & 0 & 0 & 0 & 0 & 0 & 0 & 0 & 0 & 0 & 0 & 0 & 0 & 0 \\
 1 & 0 & 0 & 1 & 0 & 0 & 0 & 0 & 0 & 0 & 0 & 0 & 1 & 0 & 0 & 1 \\
 0 & 0 & 0 & 0 & 0 & 0 & 0 & 0 & 0 & 0 & 0 & 0 & 0 & 0 & 0 & 0 \\
 0 & 0 & 0 & 0 & 0 & 0 & 0 & 0 & 0 & 0 & 0 & 0 & 0 & 0 & 0 & 0 \\
 1 & 0 & 0 & 0 & 0 & 0 & 1 & 0 & 0 & 1 & 0 & 0 & 0 & 0 & 0 & 1 \\
 0 & 0 & 0 & 0 & 0 & 0 & 0 & 0 & 0 & 0 & 0 & 0 & 0 & 0 & 0 & 0 \\
 0 & 0 & 0 & 0 & 0 & 0 & 0 & 0 & 0 & 0 & 0 & 0 & 0 & 0 & 0 & 0 \\
 1 & 0 & 0 & 0 & 0 & 0 & 1 & 0 & 0 & 1 & 0 & 0 & 0 & 0 & 0 & 1 \\
 0 & 0 & 0 & 0 & 0 & 0 & 0 & 0 & 0 & 0 & 0 & 0 & 0 & 0 & 0 & 0 \\
 0 & 0 & 0 & 0 & 0 & 0 & 0 & 0 & 0 & 0 & 0 & 0 & 0 & 0 & 0 & 0 \\
 1 & 0 & 0 & 1 & 0 & 0 & 0 & 0 & 0 & 0 & 0 & 0 & 1 & 0 & 0 & 1 \\
 0 & 0 & 0 & 0 & 0 & 0 & 0 & 0 & 0 & 0 & 0 & 0 & 0 & 0 & 0 & 0 \\
 0 & 0 & 0 & 0 & 0 & 0 & 0 & 0 & 0 & 0 & 0 & 0 & 0 & 0 & 0 & 0 \\
 4 & 0 & 0 & 1 & 0 & 0 & 1 & 0 & 0 & 1 & 0 & 0 & 1 & 0 & 0 & 4
        \end{array}\right).
    \end{equation}
    The Gilbert algorithm tested $\rho_4$ against full separability and after 7250 corrections found the following approximation of the closest separable state, with precision up to $10^{-5}$.
\begin{equation}
    \rho'_{4}\approx 10^{-5}\left(
\begin{array}{cccccccccccccccc}
 26475 & 6 & 27 & 2662 & -6 & 4 & 2669 & 0 & -38 & 2663 & 0 & -11 & 2676 & -2 & 0 & 3079 \\
 6 & 3552 & 4 & -8 & 1843 & -14 & -7 & 2320 & 23 & -3 & 6 & -2 & -14 & 2311 & -3 & 4 \\
 27 & 4 & 3547 & -6 & 1 & 3 & -3 & -3 & 1835 & -2 & 5 & 2312 & 6 & 0 & 2320 & 0 \\
 2662 & -8 & -6 & 3136 & 0 & 9 & 1550 & -16 & -11 & 1536 & -5 & -6 & 3027 & 5 & -9 & 2659 \\
 -6 & 1843 & 1 & 0 & 3574 & 0 & -5 & 2324 & 0 & -10 & -11 & 6 & 8 & 2306 & -4 & -7 \\
 4 & -14 & 3 & 9 & 0 & 3114 & -6 & -3 & 7 & -4 & -12 & -17 & 4 & 3 & -2 & 4 \\
 2669 & -7 & -3 & 1550 & -5 & -6 & 3141 & 0 & 3 & 3028 & -5 & -11 & 1532 & 1 & 11 & 2640 \\
 0 & 2320 & -3 & -16 & 2324 & -3 & 0 & 3526 & 0 & 2 & -5 & -11 & -9 & 1821 & -12 & -62 \\
 -38 & 23 & 1835 & -11 & 0 & 7 & 3 & 0 & 3559 & 2 & -2 & 2308 & 0 & -8 & 2301 & 9 \\
 2663 & -3 & -2 & 1536 & -10 & -4 & 3028 & 2 & 2 & 3134 & -4 & 0 & 1522 & -7 & 5 & 2650 \\
 0 & 6 & 5 & -5 & -11 & -12 & -5 & -5 & -2 & -4 & 3099 & 2 & -3 & -11 & 1 & -11 \\
 -11 & -2 & 2312 & -6 & 6 & -17 & -11 & -11 & 2308 & 0 & 2 & 3542 & 7 & 12 & 1830 & -4 \\
 2676 & -14 & 6 & 3027 & 8 & 4 & 1532 & -9 & 0 & 1522 & -3 & 7 & 3140 & 13 & 6 & 2652 \\
 -2 & 2311 & 0 & 5 & 2306 & 3 & 1 & 1821 & -8 & -7 & -11 & 12 & 13 & 3538 & -13 & -36 \\
 0 & -3 & 2320 & -9 & -4 & -2 & 11 & -12 & 2301 & 5 & 1 & 1830 & 6 & -13 & 3525 & -8 \\
 3079 & 4 & 0 & 2659 & -7 & 4 & 2640 & -62 & 9 & 2650 & -11 & -4 & 2652 & -36 & -8 & 26396 \\
\end{array}
\right)
\end{equation}
\end{widetext}
Table \ref{Tab1} lists the maximal values of $\bra{\phi}(\rho_4-\rho'_{4})\ket{\phi}$ for $\ket{\phi}$ belonging to different biseparablity classes.

\begin{table}
    \centering
    \begin{tabular}{|c|c|}
        \hline
         {\text Biseparability class}&{\text Maximal mean value} \\
         \hline
         \hline
         A|BCD&0.19201 \\
         \hline
         B|ACD&0.19172 \\
         \hline
         C|ABD& 0.19160\\
         \hline
         D|ABC& 0.19172\\
         \hline
         AB|CD&0.35169 \\
         \hline
         AC|BD&0.17030 \\
         \hline
         AD|BC&0.35182 \\
         \hline
    \end{tabular}
    \caption{Maximal mean values of $(\rho_4-\rho'_4)$ for different classes of pure biproduct states.}
    \label{Tab1}
\end{table}
Meanwhile, we get ${\text Tr}(\rho_4(\rho_4-\rho'_{4}))=0.35834$ and $\sqrt{{\text Tr}(\rho_4-\rho'_4)^2}=0.53969$, which translates into the found witness distance of $0.01208$. The witness operator obtained from $\rho_4-\rho'_4$ has one leading positive eigenvalue corresponding to state and. is close to $W=4(\ket{GHZ_4}\bra{GHZ_4}-\mathds{1}/16)/\sqrt{15})$, which indicates the distance to bi-separable states to be at least  $\frac{29}{12\sqrt{15}}-\frac{7}{16}\approx 0.1865$.

This scheme seems to saturate rather quickly. While being only an approximation of an optimal witness of nonexistence of BD, we consider 
\begin{equation}
    W_N=\ket{GHZ_N}\bra{GHZ_N}-\frac{1}{2}\mathds{1}_{2^N\times 2^N},
\end{equation}
which attains mean value of $\frac{1}{2}$ for the $N$-particle GHZ state and nonpositive mean values for all states with biseparable decomposition. Extending the scheme presented for $\rho_4$ we get mean value
\begin{equation}
    \mean{W_N}=-\frac{1}{2}+\frac{4}{2+2^{N/2}},
\end{equation}
which becomes negative for $N=6$. As GHZ correlations diminish exponentially, it is reasonable to accept that they are not strong enough to guarantee presence of GME.

More elaborate examples would include projected entangled product states (PEPS) \cite{verstraete}, a technique, in which ground states of large spin arrays can be found by merging singlet states.

Also, notice that the state discussed in this section are related to the optical realizations mentioned above. $\ket{\varphi_4}$ appears just in front of the beam slitter in the setup producing $\ket{GHZ_4}$, while $\rho_1(\theta)$ and $\rho_2$ are obtained from $\ket{\Psi_4}$. Ref. \cite{laskowski} showed that the four-partite visibility decreases as the pairs produced in SPDC become more distinguishable in time domain. We can use this effect to completely decohere the pairs by placing two ultrafast shutters with nonoverlapping opening times on one side of the crystal, behind the beam splitter. Subsequently, we reduce the temporal distinguishability by using narrowband spectral filters in all four outputs. Under the condition that all four measuring stations were reached by a photon, the resulting state is  $\rho_2$. If then we find one photon to be polarized in direction $(\cos\alpha,\sin\alpha)$, resulting in the state of the remaining three photons equal to $\rho_3$.
\section{Parties and particles}
To interpret the above results we can adopt a few different approaches. The first is straight-forwardly transplanted from the theory of bipartite entanglement. The key concepts are parties, who within their locations may have arbitrary large quantum systems treated uniformly and act freely upon them, but the parties are separated, The case of $\rho_3$ and $\rho_4$ can be depicted as follows.
\begin{equation}
\label{isborn}
GME=0\rightarrow{(\cdot)^{\otimes 2}}\rightarrow{\text{``local'' projections}}\rightarrow GME>0,
\end{equation}
which must hold for any potential measure of GME satisfying the discriminance requirement.  The left-most side is from the construction of $\rho_1$ and $\rho_2$, while the right-most is due to the fact that we have presented witnesses against BD of $\rho_3$ and $\rho_4$.  That means that either any measure of GME cannot be additive, even in the weakest sense (taking two copies of a state produces GME), or monotonic (GME is produced by local action). The lack of BD for $\ket{\varphi_4}$, as demonstrated by nonvanishing tripartite negativity hints towards nonadditivity of GME measures.

However, this understanding seems to be contradictory to the operational approach. Distributing another copy of a state is typically more feasible than operations on composite subsystems. On one hand, we have argued that GME opens access to more features than bipartite entanglement. on the other, here we reach the conclusion that it can be produced by simply distributing GME-free states, which questions its role as a separate resource.

Another point of view is to accept parties, but allow each of them to handle more than one particle at once. This approach seems natural for many experimental situations, e. g., quantum teleportation \cite{bennett1993} and admits BD of $\ket{\varphi_4}$, $\rho_3(\theta)$ and $\rho_4$. We are now left with two options. Either genuine multipartite entanglement was created by local operations, or it was already in a state with BD. In this paradigm, there seems to a contrast between the operational and the formal understanding of a party.

There is also a third approach possible, which is to assume that each subsystem constitutes its own party, its own location. We trivially find BDs of $\rho_1^{\otimes 2}$ and $\rho_2^{\otimes 2}$, which prevents them from being recognised as genuinely multiparticle entangled. $\rho_3$ and $\rho_4$ do not have BD, but only due to operations now understood as nonlocal, which happen to be experimentally feasible due to pairwise proximity of particles.

It remains to discuss the nature of projections $P_{A/C}$, $P_{B'}$, and $P_{B}$. While the last one contains two rows representing product states, one line of the former two is an entangled state. Still, these operators represent a component of a von Neumann measurement with {\em degenarate} results. That is to say, all states in form $(\ket{00}+e^{i\phi}\ket{11})/\sqrt{2}$ must survive projection $P_B$ intact, thus these measurements must be performed collectively on both particles. A generic state is projected onto subspace of $\{\ket{00},\ket{11}\}$, which may, and in this case does, create entanglement. This is a subtle difference between, say collectively measuring $\sigma_x\otimes\sigma_x$ on two qubits and measuring $\sigma_x$ on each of them. While the statistics turn out to be the same, the collapsed states are different. Of course, in general the resulting state may be nonentangled, but GME creation/activation protocols require that they are not.  In this way we can argue that nonentangling measurements will prevent us from creating GME.

Given a genuine three-particle entanglement measure $G3pE$, the overall three-particle entanglement 
can then be defined as
\begin{eqnarray}
   && G3pE(\rho_{AB...N})\nonumber\\
   =&&\sum_{x,y,z\in S} G3pE(\rho_{xyz}),\nonumber\\
   S=&&\{A_j,B_j,...N_j\}_{j=1}^{k}
\end{eqnarray}
where each party holds $k$ parties, and $G3E(\rho)$ is. This definition naturally provides additivity and, depending on the definition, it can be contractive under local (single-particle) operations. In this way we can define a proper measure of multi-particle entanglement. A more communication-concerned definition, would take into account the quantum link between ``macro-locations'',
\begin{eqnarray}
   && G3pE(\rho_{AB...N})\nonumber\\
   =&&\sum_{X,Y,Z}\sum_{x,y,z=1,...,k} G3pE(\rho_{X_xY_yZ_z}),\nonumber\\
   &&\{X,Y,Z\}\in\{A,...N\},\nonumber\\
   &&X\neq Y\neq Z\neq X.
\end{eqnarray}
\section{Conclusions}
We have discussed several  examples of bipartite entangled states, which after a series of free (in the sense of multipartite entanglement resource theory) operations, i.e., taking two identical copies of the same state and performing partial measurements on pairs of qubits handled by each observer, loose BD. In the orthodox paradigm of distant laboratories this signifies the presence of multipartite entanglement. Hence, genuine multipartite entanglement is not a resource with this understanding, as it it can be created from a resource free-state. We then relax the distant laboratory paradigm allowing multiple subsystems in one laboratory. This, however, implies that genuine multiparty entanglement can be produced by local actions.

By no means these examples dismiss Definition 2 as a necessary an sufficient condition for multiparticle entanglement. We want to stress, nevertheless, that in these scenarios this form of quantum correlations are created by entangling actions on two subsystems. This is in contrast to the usual activation of entanglement 
\cite{horodecki1999}. Therein, we also utilize entangling operations, but we amplify an existing quantum links. Here, with these operation we create a new form of quantum link, undoubtely absent before. 

Our examples show that the concept of genuine multipartite or multiparticle entanglement must be considered with a great caution, and with physical context in mind. Some operations, e.g. introduction of new subsystems, are may not be allowed, as they change the resource of interest itself (e.g. four- rather than three-particle entanglement).

We therefore suggest the {\em one particle per subsystem }policy. Also, it seems beneficial to focus on genuine $N$-particle entanglement rather than generic multipartite. It then becomes possible to construct a proper measure of multiparticle entanglement.

Of course, this reasoning calls for further discussion. For example, one may wonder about other requirements that need to be satisfied by entanglement measures in one particle per party paradigm. Also, additional considerations are required for multiple degrees of freedom. For example, in the case, in which a photon is entangled with one particle in polarization and in orbital angular momentum with another, one would rather see the two degrees of freedom belonging to the same subsystem.

\section{Acknowledgements}
MW akcknowledges Tomasz Paterek for a fruitful discussion. This work is supported by the Institute of Information \&
Communications Technology Planning \& Evaluation (IITP) grant funded by the
Korean government (MSIT) (No.2022-0-00463, Development of a quantum repeater in
optical fiber networks for quantum internet), by NCN
SONATA-BIS grant No. 2017/26/E/ST2/01008. Early stage of this work was supported by the Foundation for Polish
Science (IRAP project, ICTQT, Contract No. 2018/MAB/5,
co-financed by EU within Smart Growth Operational Programme)

\begin{thebibliography}{99}
\bibitem{schrodinger}E. Schr \"odinger, {\em Die Gegenw\"ortige Situation in der Quantenmechanik}, {\em Naturwissenschaften} {\bf 23}, 152 (1935).
\bibitem{bell} J. S. Bell, {\em On the Einstein-Podolsky-Rosen Paradox},
{\em Physics Physique Fizika} {\bf 1}, 195 (1964).
\bibitem{ekert} A. K. Ekert, {\em Quantum Cryptography Based on Bell’s Theorem}, {\em Physical Review Letters} {\bf 67}, 661 (1991).
\bibitem{peres} A. Peres, {\em Separability Criterion for Density Matrices},
{\em Physical Review Letters} {\bf 77}, 1413 (1996).
\bibitem{horodecki1996} R. Horodecki, M. Horodecki, and P. Horodecki, {\em Teleportation, Bell’s Inequalities and  Inseparability}, {\em Physics Letters A} {\bf 222}, 21 (1996).
\bibitem{horodecki1997} P. Horodecki, {\em Separability Criterion and Inseparable
Mixed States with Positive Partial Transposition}, {\em Physics
Letters A} {\bf 232}, 333 (1997).
\bibitem{jamiolkowski} A. Jamio\l{}kowski, {\em Linear Transformations which Preserve
Trace and Positive Semidefiniteness of Operators}, {\em Reports
on Mathematical Physics} {\bf 3}, 275 (1972).
\bibitem{choi} M.-D. Choi, {\em Completely positive linear maps on complex matrices}, {\em Linear Algebra and its Applications} {\bf 10}, 285 (1975).
\bibitem{terhal} B. M. Terhal, {\em Bell Inequalities and the Separability Criterion}, {\em Physics Letters A} {\bf 271}, 319 (2000).
\bibitem{bennett1996} C. H. Bennett, D. P. DiVincenzo, J. A. Smolin, and W. K. Wootters, {\em Mixed-state Entanglement and Quantum Error Correction}, {\em Physical Review A} {\bf 54}, 3824 (1996).
\bibitem{vedral} V. Vedral and M. B. Plenio, {\em Entanglement Measures and Purification Procedures}, {\em Physical Review A} {\bf 57}, 1619 (1998).
\bibitem{eisert2003} J. Eisert, K. Audenaert, and M. Plenio, {\em Remarks on Entanglement Measures and non-local state distinguishability}, {\em Journal of Physics A: Mathematical and General} {\bf 36},
5605 (2003).
\bibitem{chen} K. Chen and L.-A. Wu, {\em Test for Entanglement Using Physically Observable Witness Operators and Positive Maps}, {\em Physical Review A} {\bf 69}, 022312 (2004).
\bibitem{hiesmayr} B. C. Hiesmayr, M. Huber, and P. Krammer, {\em Two Computable Sets of Multipartite Entanglement Measures}, {\em Physical Review A} {\bf 79}, 062308 (2009).
\bibitem{greenberger} D. M. Greenberger, M. A. Horne, and A. Zeilinger, {\em Going
Beyond Bell’s Theorem}, in {\em Bell’s theorem, quantum theory and conceptions of the universe} (Springer, 1989) pp. 69–72.
\bibitem{guehne} O. G\"uhne, G. T\ 'oth, and H. J. Briegel, {\em Multipartite Entanglement in Spin Chains}, {\em New Journal of Physics} {\bf 7}, 229 (2005).
\bibitem{steane} A. M. Steane, {\em Error Correcting Codes in Quantum Theory}, {\em Physical Review Letters} {\bf 77}, 793 (1996).
\bibitem{yamasaki} H. Yamasaki, S. Morelli, M. Miethlinger, J. Bavaresco,
N. Friis, and M. Huber, {\em Activation of Genuine Multipartite Entanglement: Beyond the Single-copy Paradigm of
Entanglement Characterisation}, {\em Quantum} {\bf 6}, 695 (2022).
\bibitem{bengtsson} I. Bengtsson and K. \.Zyczkowski, {\em Geometry of Quantum
States: an Introduction to Quantum Entanglement}, (Cam-
bridge university press, 2017).
\bibitem{zyczkowski} K.  \.Zyczkowski, P. Horodecki, A. Sanpera, and M. Lewen-
stein, {\em Volume of the Set of Separable States}, {\em Physical Review A} {\bf 58}, 883 (1998).
\bibitem{eisert1999} J. Eisert and M. B. Plenio, {\em A Comparison of Entanglement
Measures}, {\em Journal of Modern Optics} {\bf 46}, 145 (1999).
\bibitem{henderson} L. Henderson and V. Vedral, {\em Information, Relative Entropy of Entanglement, and Irreversibility}, {\em Physical Review Letters} {\bf 84}, 2263 (2000).
\bibitem{witte} C. Witte and M. Trucks, {\em A Aew Entanglement Measure
Induced by the Hilbert–Schmidt Norm}, {\em Physics Letters A}
257, {\bf 14} (1999).
\bibitem{pandya} P. Pandya, O. Sakarya, and M. Wie\'sniak, {\em Hilbert-Schmidt Distance and Entanglement witnessing}, {\em Physical
Review A} {\bf 102}, 012409 (2020).
\bibitem{hu} Y. Hu, Y.-C. Liu, and J. Shang, {\em Algorithm for Evaluating Distance-Based Entanglement Measures}, {\em Chinese Physics B} {\bf 32}, 080307 (2023).
\bibitem{ma2023} M. Ma, Y. Li, and J. Shang, {\em Multipartite Entanglement Measures: a Review} (2023), arXiv:2309.09459 [quant-ph].
\bibitem{ma2011} Z.-H. Ma, Z.-H. Chen, J.-L. Chen, C. Spengler,
A. Gabriel, and M. Huber, {\em Measure of Genuine Multipartite Entanglement with Computable Lower Bounds}, {\em Physical Review A} {\bf 83}, 062325 (2011).
\bibitem{mintert} F. Mintert, A. R. Carvalho, M. Ku\'s, and A. Buchleitner, {\em Measures and Dynamics of Entangled States}, {\em Physics
Reports} {\bf 415}, 207 (2005).
\bibitem{gilbert} E. G. Gilbert, {\em An Iterative Procedure for Computing the Minimum of a Quadratic Form on a Convex Set}, {\em SIAM Journal on Control} {\bf 4}, 61 (1966).
\bibitem{brandao} F. G. Brand\~{a}o, {\em Quantifying Entanglement with Witness Operators}, {\em Physical Review A} {\bf 72}, 022310 (2005).
\bibitem{louisell} W. Louisell, A. Yariv, and A. Siegman, {\em Quantum Fluctuations and Noise in Parametric Processes. I.}, {\em Physical Review} {\bf 124}, 1646 (1961).
\bibitem{pan} J.-W. Pan, M. Daniell, S. Gasparoni, G. Weihs, and A. Zeilinger, {\em Experimental demonstration of four-photon entanglement and high-fidelity teleportation}, {\em Physical Review Letters} {\bf 86}, 4435 (2001).
\bibitem{gaertner} S. Gaertner, M. Bourennane, M. Eibl, C. Kurtsiefer, and H. Weinfurter, {\em High-fidelity source of four-photon entanglement}, {\em Applied Physics B} {\bf 77}, 803 (2003).
\bibitem{sabin} C. Sab\'in and G. Garc \'ia-Alcaine, {\em A Classification of Entanglement in Three-qubit Systems}, {\em The European Physical
Journal D} {\bf 48}, 435 (2008).
\bibitem{verstraete} F. Verstraete and J. I. Cirac, {\em Renormalization Algorithms for Quantum Many-body Systems in Two and Higher Dimensions} (2004), arXiv:cond-mat/0407066
[cond-mat.str-el].
\bibitem{laskowski} W. Laskowski, M. Wie\'sniak, M. \.Zukowski, M. Bourennane, and H. Weinfurter, {\em Interference contrast in multi-source few-photon optics}, {\em Journal of Physics B: Atomic, Molecular and Optical Physics} {\bf 42}, 114004 (2009).
\bibitem{bennett1993} C. H. Bennett, G. Brassard, C. Cr\'epeau, R. Jozsa, A. Peres, and W. K. Wootters, {\em Teleporting an Unknown
Quantum State via Dual Classical and Einstein-Podolsky-
Rosen Channels}, {\em Physical Review Letters} {\bf 70}, 1895 (1993).
\bibitem{horodecki1999} P. Horodecki, M. Horodecki, and R. Horodecki, {\em Bound Entanglement Can Be Activated}, {\em Physical Review Letters} {\bf 82}, 1056 (1999).
\end{thebibliography}

\end{document}